# Bifurcations and Complete Chaos for the Diamagnetic Kepler Problem

Kai T. Hansen[*]
Fakultät für Physik, Universität Freiburg
Hermann-Herder-Strasse 3, D-79104 Freiburg, Germany
e-mail: k.t.hansen@fys.uio.no

February 6, 1995

**Abstract**

We describe the structure of bifurcations in the unbounded classical Diamagnetic Kepler problem. We conjecture that this system does not have any stable orbits and that the non-wandering set is described by a complete trinary symbolic dynamics for scaled energies larger then $\epsilon_c = 0.328782\ldots$.

The problem of a charged particle moving in a combined attracting electrical Coloumb field and a strong homogeneous magnetic field is a simple system showing complicated chaotic behavior. The corresponding quantum mechanical problem is a Hydrogen atom in a strong magnetic field. This system is called the *Diamagnetic Kepler problem* (DKP) and in the last decade this problem has been much investigated both classicly and quantum mechanically in theoretical and experimental studies. Review articles by Friedrich and Wintgen[1] and Hasegawa et.al.[2] yield good introductions to this problem. One interesting result of this research is an astonishing relation between the classical and the quantum mechanical systems, where the quantum mechanical energy levels analyzed in a proper way reflect the periodic orbits existing in the classical system. Most of these results have been on the bounded system with negative energy such that the electron cannot escape. We report here some numerical results concerning the classical orbits in the DKP with a positive energy, where quantum mechanically the atom can ionize. We also give some remarks on the implications for the quantum mechanical systems from the results obtained here.

The Diamagnetic Kepler problem with zero angular momentum is given by the Hamiltonian

$$H = p^2/2m_e - e^2/r + \frac{1}{2}m_e\omega^2(x^2+y^2) \tag{1}$$

Using suitably scaled variables and semi-parabolic coordinates this problem can be turned into the study of the Hamiltonian[1, 2]

$$h = \frac{1}{2}p_\nu^2 + \frac{1}{2}p_\mu^2 - \epsilon(\nu^2+\mu^2) + \frac{1}{8}\nu^2\mu^2(\nu^2+\mu^2) \equiv 2 \tag{2}$$

at the fixed "pseudo energy" 2 and with a scaled energy $\epsilon = E\gamma^{-2/3}$ depending on both the energy, $E$, and the strength of the magnetic field, $\gamma$. We consider the scaled energy $\epsilon$ to be a parameter when we investigate the classical trajectories.

The Hamiltonian (2) has a rather complicated bifurcation structure for $\epsilon < 0$ and there is not any known symbolic dynamics description of the orbits. For $\epsilon \geq 0$ it has been proposed by Eckhardt and Wintgen[3] that the orbits may be described by the same symbols as the 4-disk system [4] and that this 4-disk symbolic dynamics for periodic orbits (cycles) can be obtained in DKP by finding self conjugate points. The 4-disk system consists of a point particle moving freely on a 2-dim. plane and bouncing elastically off the boundaries of 4 disks in the plane. The 4 disks have radius 1 places with the disc centers at the four points $(\pm r/2, \pm r/2)$ int the $(\nu, \mu)$-plane. The center-center distance $r$ is the parameter for the 4-disk problem.

For the Hamiltonian system with the smooth potential $V(x, y) = (xy)^{2/a}$ it has been argued that the bifurcations are very similar in this smooth potential and in the 4 disk billiard[7, 8]. This seems to be true also for the DKP with positive $\epsilon$ and this yields a method to determine stable orbits and to find the parameter value where the repellor is a complete Cantor set. We can then show that there are no elliptic orbits above a critical energy $\epsilon_c$ which has to be determined numerically.

In a dispersing billiard system the singular orbits which define the pruning front [9, 10, 11] are the orbits tangent to the boundary. The pruning front determines which orbits that are admissible in the system and all bifurcations take place on this pruning front. In the 4-disk billiard we obtain a symbolic dynamics description by enumerating the disks clockwise and the symbolic dynamics of a trajectory is given by a string $\ldots s_{-1}s_0s_1s_2\ldots$ where $s_t$ is the label of the disk the particle bounces off at the integer time $t$. Other alphabets can be defined from this basic alphabet [4, 11]. The first orbits which become forbidden in the 4-disk system as the distance between the disk centers decreases to a critical distance $r_c = 2.2046\ldots$, are the heteroclinic orbits $\overline{14}\,\overline{21}$ and $\overline{14}\,1\,\overline{21}$ (a line above a string denotes an infinite repetition of this string $\overline{14}\,\overline{21} = \cdots 141414212121\cdots$) and orbits symmetric to these [10, 11]. These orbits are the corners of the 2-dimensional Cantor set of non-wandering[5] (not escaping) orbits in the phase space and are the first to become non-admissible in a symmetric system. We show below that these heteroclinic orbits also are the corners of the non-wandering Cantor set for the complete chaotic DKP.

The idea that singular orbits determine the set of admissible orbits can be carried over from the hard billiard systems to the soft potential systems. One then has to determine the orbits in the smooth system which correspond to the tangential orbits in the billiard. A second way to view the analogy between hard and soft systems is to draw the manifold structures in a Poincaré map. This yields a simpler and maybe more familiar picture, but the symbolic dynamics and physical interpretation are more difficult to get from this point of view. We will here discuss both approaches to the problem.

There does not exist a steep wall in the smooth system as it does in the billiard system and consequently no direct way to find a "tangent" orbit, but the singular orbits in the billiards have another property to which we can find an analog to in smooth systems and we will apply this to determine the singular orbits. Assume that there exists a singular orbit tangent to the boundary in the 4 disk system. Then choose a second orbit close to the singular orbit at the tangent point but a finite distance $\delta$ away from the boundary and such that this second orbit does bounce off different disks than the first orbit after some time. This second orbit will then have a different symbolic description from the singular orbit. Then there will also exist a third orbit bouncing off the boundary near the tangent point of the first orbit with some finite angle $\phi$ with a symbolic description identical to the second orbit both in the future and in the past. We can define the singular orbit as an orbit with extremum future and past trajectories since the two different neighbor trajectories have identical symbolic dynamics in past and future. In a well ordered symbolic alphabet [11] this yields a extremum symbolic value and in the configuration space the two neighbor orbits moves on the same side of the singular trajectory. This new definition of singular orbits as extremum orbits can be applied also in the case of a smooth potential.

The set of all singular orbits is a one-parameter family of orbits in the 4-disk system and this seems to be true also for the smooth Hamiltonian (2) in the DKP.

In the following we for simplicity mostly discuss the singular orbit normal to the diagonal $\mu = \nu$. It is clear from the symmetry of the problem that there will exist such a singular orbit and as for the 4-disk system this singular orbit becomes the heteroclinic orbit $\overline{14}\,\overline{21}$ at the critical parameter value $\epsilon_c$. Then there is no pruning for $\epsilon > \epsilon_c$. The singular orbit on the $\nu_0 = \mu_0$ line can be found numerically by scanning a one dimensional starting point $\nu_0 = \mu_0$ keeping $p_{\nu_0} = -p_{\mu_0}$ while for other non-symmetric singular orbits one has to scan both the starting points and the direction of the velocity.

Fig. 1 shows a segment of the singular orbit for the scaled energy $\epsilon = 0$ from the starting point $\nu_0 = \mu_0 = 0.816$ together with two neighboring orbits starting parallel with this at $\nu_0 = \mu_0 = 0.716$ (dashed curve) and at $\nu_0 = \mu_0 = 0.916$ (dotted curve). Both the two neighbor trajectories are on the same side of the singular trajectory after some transient time both in the future and in the past. There are no trajectories on the other (the forbidden) side of the singular trajectory both in the past and in the future. There may be orbits on this forbidden side in the past *or* in the future, but not both in the past and in the future.

The orbit starting at $\nu_0 = \mu_0 = 0.716$ corresponds to the orbit in the billiard system which passes the boundary of disk no. 1 at some distance $\delta$, while the orbit starting at $\nu_0 = \mu_0 = 0.916$ corresponds to the billiard orbit bouncing off the boundary of disk no. 1 with some angle $\phi$. By drawing the orbits in configuration space we can determine the singular orbit and we find that the singular orbit for $\epsilon = 0$ starts at $\nu_0 = \mu_0 = 0.8166179\ldots$.

The singular orbit in Fig. 1 shows that there cannot exist any orbit described by a symbol string containing a substring _414212_ or _4141212_ for $\epsilon = 0$. This orbit would be on the forbidden side of the singular trajectory. One forbidden periodic orbit is then $\overline{41412121}$. Since the symmetric singular orbit for $\epsilon = 0$ numerically appears to be non-periodic there is an infinite number of forbidden symmetric sub-strings of increasing length determined by this singular orbit.

To determine all forbidden sub-strings one also has to determine the non-symmetric singular orbits. One example of a non-symmetric singular orbit is drawn in Fig. 2. We have here numerically determined a trajectory with extremum future and past compared to orbits starting parallel to it. This orbit starts at $\nu_0 = 0.56166$, $\mu_0 = 1.56166$ and with $p_{\nu_0}/p_{\mu_0} = -\tan(0.605)$. From Fig. 2 we can read off that for example the orbits containing a substing _1421212_ or _14121212_ are forbidden. This is a slightly different rule from that we got from Fig. 1. All the rules obtained from all singular orbits yield all rules which determine the forbidden orbits in the system.

As $\epsilon$ increases the singular orbits will change and for some values of $\epsilon$ a singular orbit may return close to the starting point with a similar direction. When the singular orbit returns it implies that we get a stable periodic orbit. Since orbits near the singular orbit do not diverge but approach each other around the starting point of the singular orbit this gives rise to stability for periodic orbits. One example of this is given in Fig. 3 for the scaled energy $\epsilon = 0.1113$ where the singular orbit is found at $\mu_0 = 0.8617$. The stable cycle existing here is $\overline{41412121}$. This cycle we found above to be inadmissible at $\epsilon = 0$ and it will be admissible for all $\epsilon > 0.1113$. The cycle is in the center of a stable island and it is not directly a forbidden and legal side of the singular orbit here, but the admissibility is determined by the rotation number of the elliptic orbits. The nearby orbits drawn in Fig. 3 start outside the stable island.

Inserted in Fig. 3 is a Poincaré map showing $(\nu, p_\nu)$ when the trajectory crosses the $\nu = \mu$ line for 5 different starting points close to the singular trajectory $(0.855 < \nu < 0.869, 1.4354 < p_\nu < 1.4375)$. We find here the usual Hamiltonian structure of KAM-curves and resonances. We conjecture that all the periodic and quasiperiodic orbits within the stable island can be given a unique symbolic coding by adiabatically following the trajectories into the completely chaotic region $\epsilon > \epsilon_c$, and that they belong to one bifurcation family ot the 4-disk system, se ref. [8]. The symbolic dynamics of these orbits are given as a string $\cdots 414 s_{-1} 212 s_0 414 s_1 212 \cdots$ where $s_t$ is the symbol 1 or no symbol.

All the cycles symmetric across the $\mu = \nu$ line are stable when they are on the symmetric singular orbit and hyperbolic otherwise.

We will now try to find the parameter value $\epsilon_c$ where the singular orbit escapes directly to $\nu \to \infty$ in the future direction and to $\mu \to \infty$ in the past direction without ever returning. It is however not trivial to decide when a trajectory leaves to infinity in this system because for an arbitrary large value of $\nu$ (or of $\mu$) there exist orbits coming from and returning to the $\nu = 0$ line ($\mu = 0$ line). One example of this is an orbit crossing the $\nu$-axis with $p_\nu = 0$ for a large value of $\nu$. This trajectory will cross the $\nu = 0$ line both in the future and in the past and may be a not escaping trajectory.

Numerically it turns out that it is possible to distinguish escaping and not escaping trajectories by plotting the Poincaré map $(\mu, p_\mu)$ for each crossing of the $\mu$-axis with $p_\mu > 0$ (correspondingly for $\nu$). Escaping trajectories yield points approximately along or above a straight line $p_\mu = a\mu + b$ with $a > 0$. Returning trajectories will yield points along a oval shaped curve in the Poincaré map $(\mu, p_\mu)$. We can then distinguish between escaping and not escaping trajectories by following the trajectory a reasonably short time.

Fig. 4 shows the Poincaré map for three orbits starting at $\mu = \nu = 0.75$ ($\diamond$), 0.9 (+) and 1.05 ($\square$), $p_{\nu_0} = -p_{\mu_0}$, for the scaled energy $\epsilon = 0.335$. The trajectory from 0.9 escapes directly while the other two return to the $\mu = 0$-line (but may escape at some later time). The escaping trajectory is drawn in configuration space $(\nu, \mu)$ in Fig. 5. The orbit simply comes in from $\mu = \infty$ and continues out to $\nu = \infty$. When the symmetric singular orbit escapes directly there is no pruning left in the system and we will have a complete Cantor set repellor. Other non-symmetric singular orbits will for some $\epsilon < \epsilon_c$ escape to infinity either in the future or in the past. There will not be any stable cycles since stable cycles only appear near a non-escaping singular orbit.

There is one orbit missing from the Cantor set and this is the cycle $\overline{12}$ which is only approached in the $\nu \to \infty$ limit. Orbits approaching this cycle, for example $\overline{34(12)^n}$; $n \to \infty$, are then not bounded by a maximum value of $\nu$, but they are unstable.

Numerically we determine the critical parameter value to be $\epsilon_c = 0.328782\ldots$.

A different way to discuss this problem is to draw the manifold structures in a Poincaré plane. In this way we can visualize how the stable and the unstable manifolds cross each other with a finite angle for $\epsilon > \epsilon_c$. When all stable and unstable manifolds cross with a finite angle (no tangencies), then the system is hyperbolic. In Fig. 6 (a) we have drawn all starting points in a Poincaré plane not escaped after some fixed integration time $T$ for the scaled energy $\epsilon = 0.35$. We have chosen the Poincaré plane where the trajectory crosses the $\nu = \mu$ line from $\nu < \mu$ to $\nu > \mu$ with the coordinates $(\mu, \phi)$, where $\phi$ is the angle of the trajectory with the $\nu$-axis; $p_\mu / p_\nu = \tan \phi$. This Poincaré plane is finite with $-3\pi/4 < \phi < \pi/4$ and $-\mu_{\max} < \mu < \mu_{\max}$ where $\mu_{\max}(\epsilon)$ is found analytically from (2). To determine if a trajectory has escaped we use the idea from Fig. 4 that an orbit escapes if the value of $p_\mu$ ($p_\nu$) increases from one to the next crossing of the $\mu$-axis ($\nu$-axis). The stable manifolds, $W^s$, are included in the set of not escaping trajectories, and for a sufficiently large $T$ this yields a good picture of $W^s$. All points on the border curve, $\phi = -3\pi/4$, $\phi = \pi/4$, or $\mu = \pm\mu_{\max}$ in the Poincaré map, correspond to the hyperbolic period two cycle $\overline{13}$. The structure close to the border can be mapped into the structure in the middle of the Poincaré plane (close to the period two cycle $\overline{24}$) by e.g. $\nu \to -\nu$. Below we also draw the manifolds in a Poincaré map that does not have this strange border but which on the other hand is infinite.

The unstable manifolds $W^u$ are symmetric to $W^s$ with respect to the line $\phi = -\pi/4$. Both $W^u$ and $W^s$ are drawn in Fig. 6 (b) and the crossing points between the $W^u$ and $W^s$ curves yield the non-wandering set. We find here that all the crossings between the curves of $W^u$ and $W^s$ have a finite angle and consequently: all orbits are hyperbolic. In Fig. 6 it is simple to recognize a three-interval Cantor set structure of the manifolds. It is slightly more complicated to recognice this at the edges because the border curve corresponds to one unstable cycle, and we have here only one half of the Cantor set. This only depends on our choice of Poincaré map and does not give any real problems.

These figures give a similar structure of manifolds as the manifold structure one can find for a 4-disk system. The stable and the unstable manifolds for the 4-disk system in a corresponding Poincaré map are drawn in Fig. 7. In Fig. 7 (a) both the stable and the unstable manifolds are drawn for the disk-disk distance $r = 2.35$, $-3\pi/4 < \phi < \pi/4$, $1/\sqrt{2} - r/2 < \mu < r/2 - 1/\sqrt{2}$. This yields nice trinary Cantor sets as crossing points between $W^u$ and $W^s$ in the same way as we found it in Fig. 6 (b). The main structure is exactly the same in the billiard system, Fig. 7 (a), and in the smooth system, Fig. 6 (b). The important difference is that in DKP all folds smoothly connect for large values of $|\mu|$ while for the 4-disk system the manifolds are discontinuous at the line $\mu = \mu_{\max}$. This discontinuous jump from $\phi$ to $\pi/2 - \phi$ for the folds of the 4-disk system is loosely speaking for DKP glued together and the folds smoothly connect at the line $\phi = -\pi/4$. The two drawings yield the same number of crossings between $W^u$ and $W^s$.

We can also compare the pruned systems for the DKP ($\epsilon < \epsilon_c$) and the 4-disk system ($r < r_c$) in this Poincaré plane. In Fig. 7 (b) the stable manifold $W^s$ for the 4-disk system is drawn for the disk distance $r = 2.06$. We have here drawn a quarter of the full phase space $-\pi/4 < \phi < \pi/4$, $0 < \mu < r/2 - 1/\sqrt{2}$. We find that the largest white area which corresponds to orbits escaping without ever returning, does not any more reach the $\phi = -\pi/4$ line as it does in Fig. 7 (a). Some crossings between the stable and the unstable manifolds existing for $r = 2.35$ are here lost, and then also an infinite number of periodic orbits are inadmissible. A sketch of the manifold structure are given in Fig. 8 (a). The points where a cusp of the stable manifold touch a fold of the unstable manifold are marked with circles. These singular points are exactly the images of the tangent orbits. Moving the disks slightly closer to each other makes the homoclinic orbit at this touching point inadmissible. This is the analog in a billiard to the point called a homoclinic tangency for smooth dynamical systems. For simplicity we also denote this point a homoclinic tangency. One important remark here is that the homoclinic tangencies indicated in Fig. 8 (a) are the primary tangencies and all other tangencies in other part of the phase space are images of these. The different primary tangencies are associated with different tangent orbits in the billiard. There is a one to one correspondence between a primary homoclinic tangency and singular orbits in the billiard. The homoclinic tangencies can be used to define the pruning front.

In Fig. 9 manifolds are drawn for the Diamagnetic Kepler problem with $\epsilon = 0.2 < \epsilon_c$. In Fig. 9 (a) both the stable and the unstable manifolds are drawn while in Fig. 9 (b) only the stable manifolds are drawn in one quarter of the phase space $0 < \mu < \mu_{\max}$, $-\pi/4 < \phi < \pi/4$. Comparing Fig. 9 (b) with Fig. 7 (b) we find a very similar structure. The main difference is that the mainifolds in Fig. 9 (b) are smooth while in Fig. 7 (b) the bending points are sharp. For the DKP we have homoclinic tangencies where the unstable and the stable manifolds are tangents and for an infinitesimal smaller parameter $\epsilon$ these homoclinic orbits are inadmissible. The structure of the homoclinic tangencies is sketched in Fig. 8 (b). As for the 4-disk system we have a one parameter family of primary homoclinic tangencies each associated with a singular orbit. The situation in the soft potential is slightly more complicated because a stable orbit creates a stable island in the neighborhood of a homoclinic tangency. One example of such an island is the orbit in Fig. 3. One way to numerically determine the pruning front is to draw the unstable and the stable manifolds and determine the set of primary homoclinic tangencies. This can also be done for the closed system ($\epsilon = 0$) by using the manifolds of a hyperbolic periodic orbit.

To further illustrate the hyperbolicity for this system, we draw the manifolds in a different Poincaré map where the trajectory crosses the $\mu$-axix with $\nu$ changing from negative to positive. The manifolds in Fig. 9 (b) for the DKP at $\epsilon = 0.20$ are drawn in Fig. 10 (a). Here there is a number of homoclinic tangencies giving rise to pruning and stable orbits. The manifolds in Fig. 6 (b) for the hyperbolic map at $\epsilon = 0.35$ are drawn in Fig. 10 (b). Here we find that the manifolds always cross with a finite angle and there are no tangencies. In Fig. 10 (c) the same manifolds are drawn for a large scaled energy, $\epsilon = 2.0$. Here we find that the same Cantor set of crossings between the stable and the unstable manifolds but the Cantor set is thinner. It is not created any new non-escaping orbits when $\epsilon$ increases because the manifolds only moves closer to the $\phi = 0$ and $\phi = \pi$ lines. It is reasonable to conjecture that no new orbits are created as $\epsilon$ increases further and the non-wandering set of the DKP has the same topological structure for all $\epsilon > \epsilon_c$.

An advantage of this Poincaré map is that we avoid the complicated border orbit we had in Figs. 6 and 9, and it is easy to see that no new orbits are created as $\epsilon$ increases. The disadvantage is that the plane is unbounded and as $\mu$ increases we get more and more copies of the foliations. Each new copy only has one more bounce than the presiding one and does not yield any new information.

In Fig. 11 equipotential lines are drawn at $\epsilon_c$. In addition to the hyperbola like curves also existing at $\epsilon = 0$, there are saddle points and a "hill" at the origin. These latter structures turn out not to give rise to any new singular orbits and stable cycles. The potential at the origin is 0 and trajectories cannot turn there. The hill structure only adds instability to the Cantor set repellor.

In quantum mechanical calculations [12, 13, 14] one has found resonances and some with a very narrow width. In quantum mechanical calculations these appear because of cancellation of some matrix elements. Semiclassically these can be associated with stable orbits like the one in Fig. 3. When we have a pure hyperbolic classical system ($\epsilon > \epsilon_c$) we can predict that a quantum mechanical cross section (for $\epsilon$ constant) does not have any narrow width resonances. All the other resonances can be calculated by periodic orbit expansion of the unstable periodic orbits and work on this continues.

The DKP with negative scaled energy has a more complicated structure and more mechanisms giving stable orbits. For small negative energies the pruning front orbits discussed above still exist as a part of the complete pruning mechanism.

From the results presented above it seems like the Hamiltonian (2) with $\epsilon > \epsilon_c$ is a completely chaotic scattering smooth system, without any singularities and where the set of all periodic orbits (and the non-wandering set) is unbounded in the configuration space. To the authors knowledge there are no earlier examples of this in the literature.

The author wish to express his gratitude to Dieter Wintgen for the few interesting discussions time allowed. The author thanks friends in Copenhagen and Freiburg for good discussions. The author is grateful to the the Alexander von Humboldt foundation for financial support.

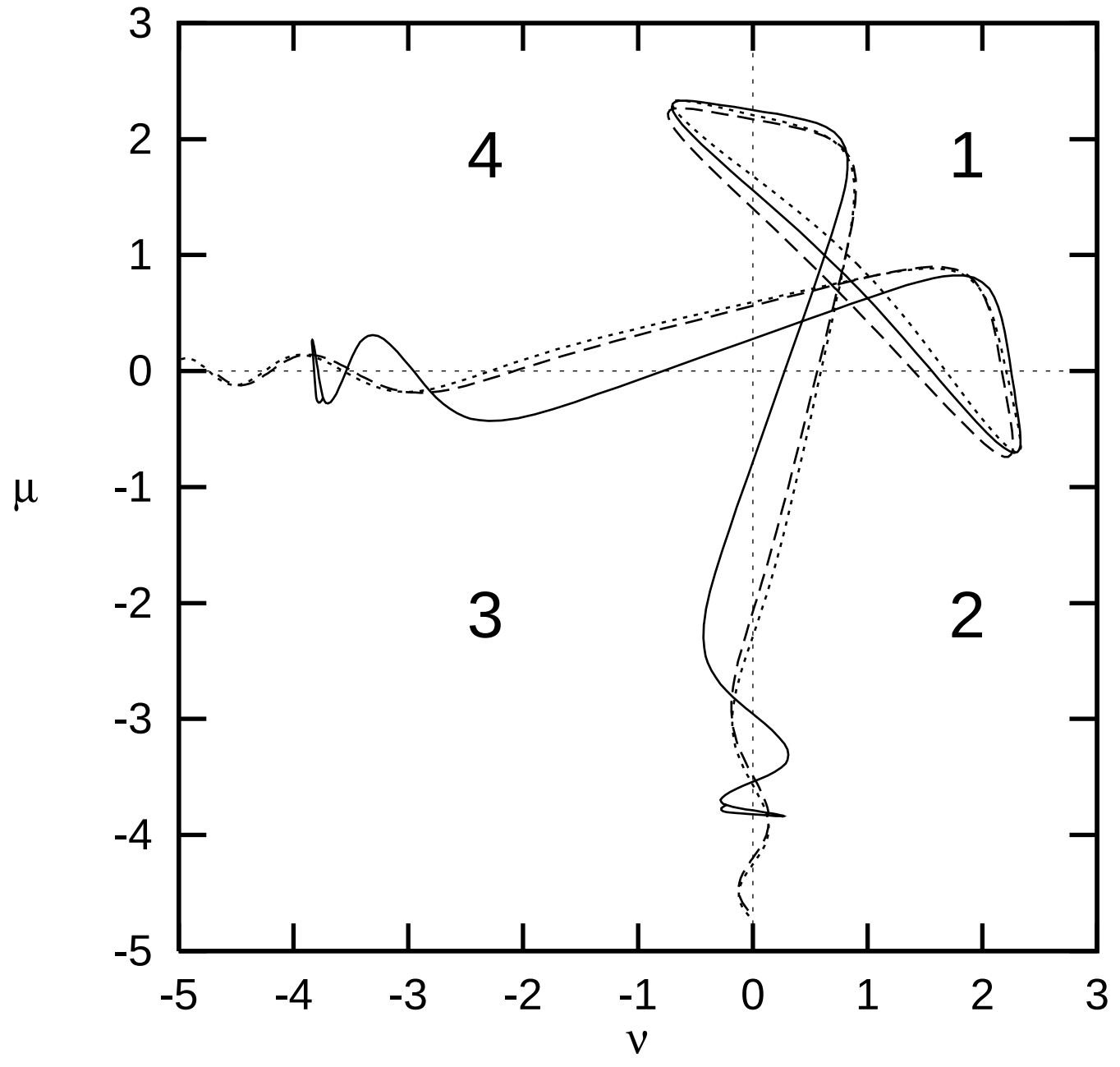

Figure 1: The symmetric singular orbit and two neighbor orbits for $\epsilon = 0$ in the diamagnetic Kepler problem (DKP). The numbers indicate the symbols of the 4-disk system.

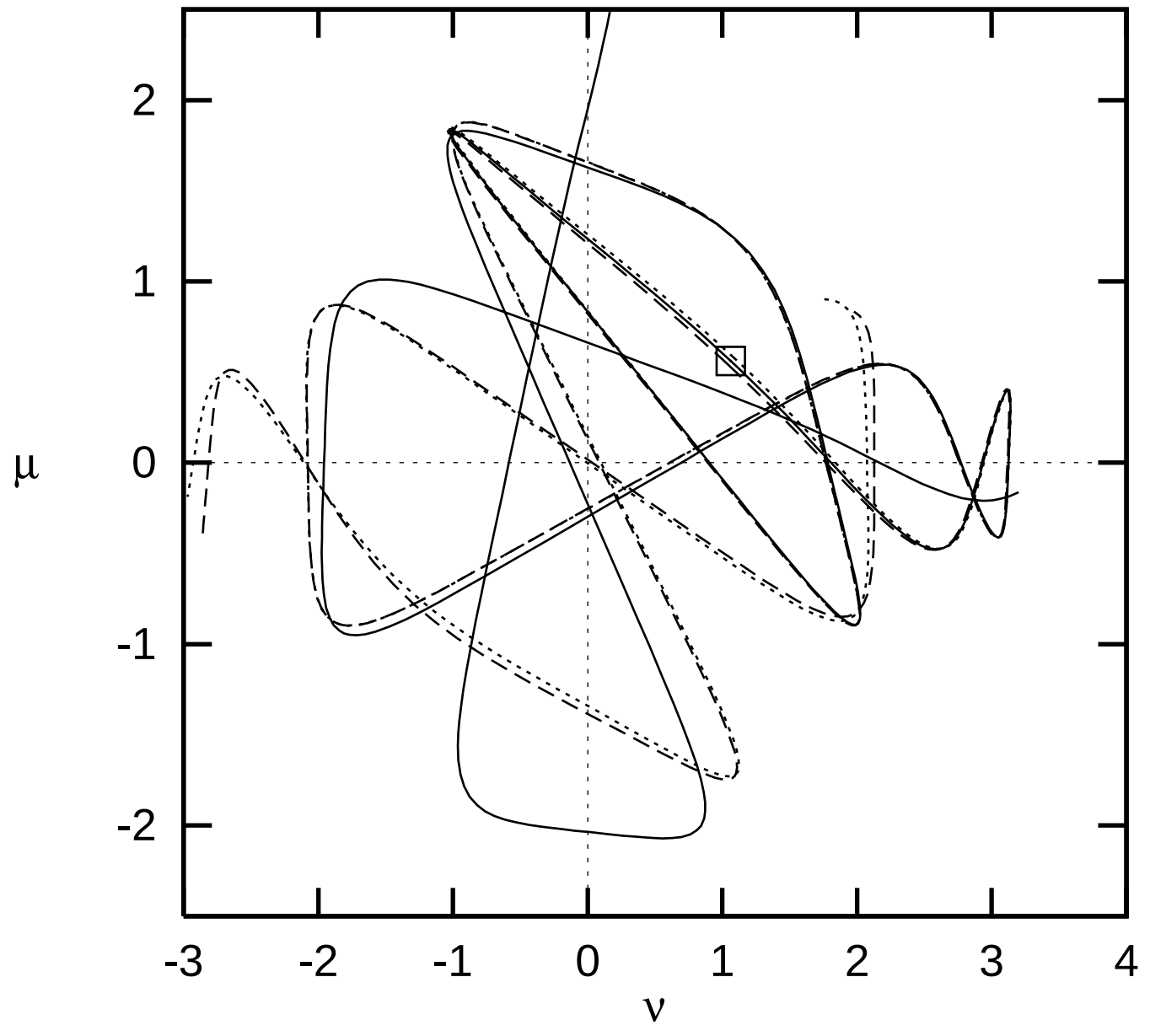

Figure 2: A non-symmetric singular orbit and two neighbor orbits for $\epsilon = 0$.

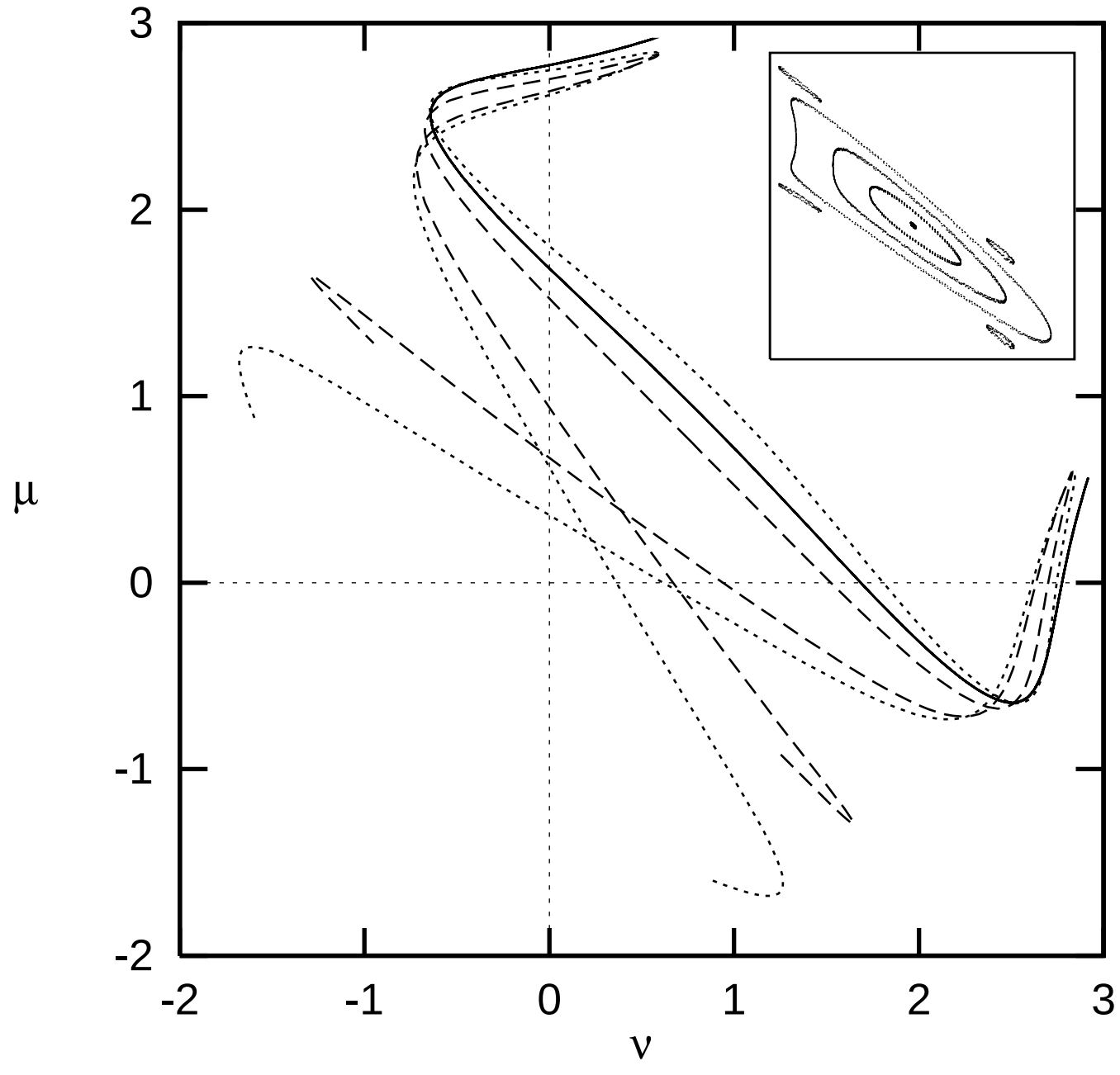

Figure 3: A singular orbit and two neighbor orbits for $\epsilon = 0.1113$ where the singular orbit is a stable periodic orbit. Inserted is a Poincaré map $(\mu, \phi)$ for $\nu = \mu$ showing KAM-curves.

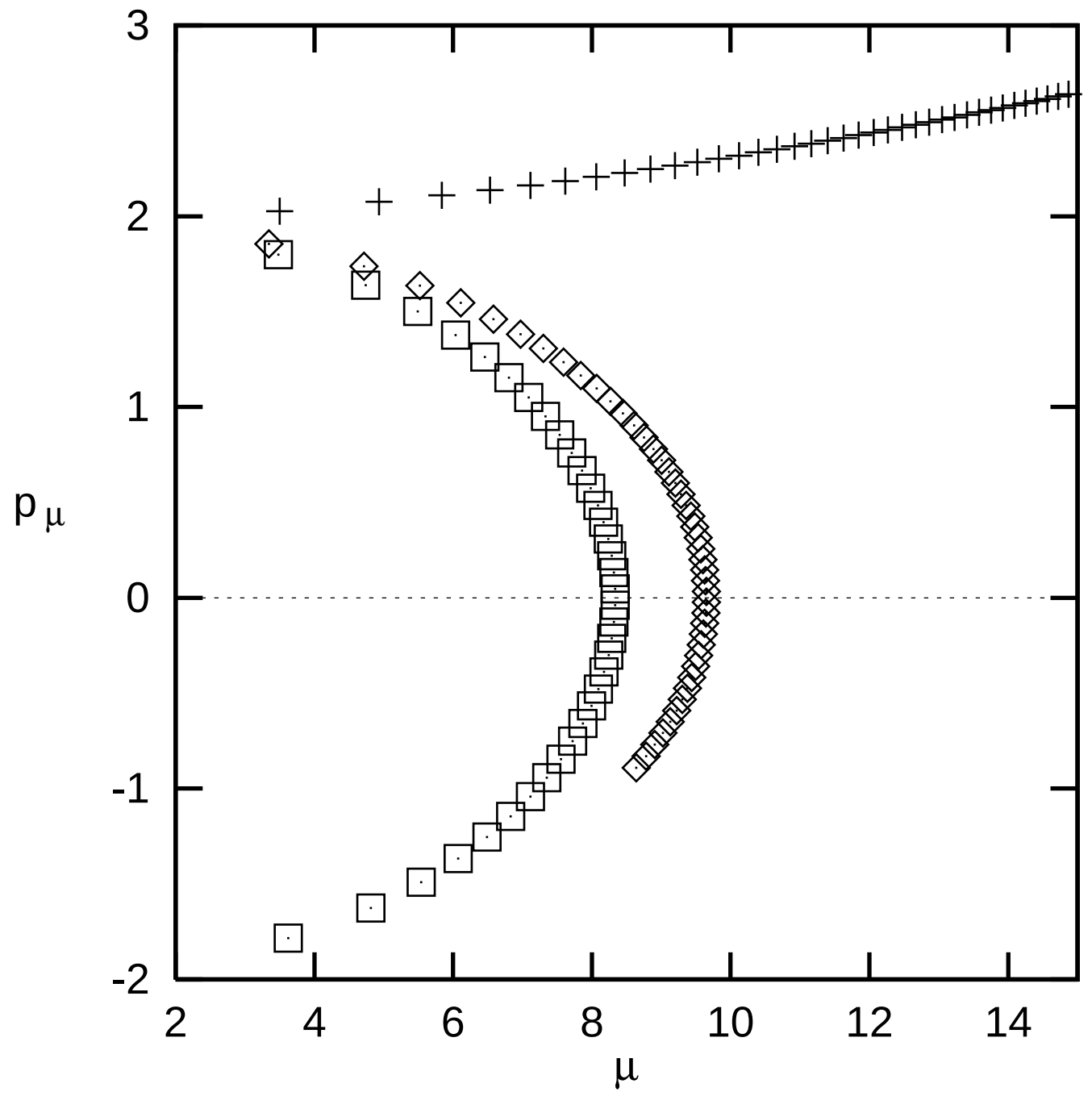

Figure 4: The Poincaré map $(\mu, p_\nu)$ for $\nu = 0$ of three trajectories from $\nu_0 = \mu_0, p_{\nu_0} = -p_{\mu_0}$ showing that there is one trajectory escaping directly yielding a completely chaotic system.

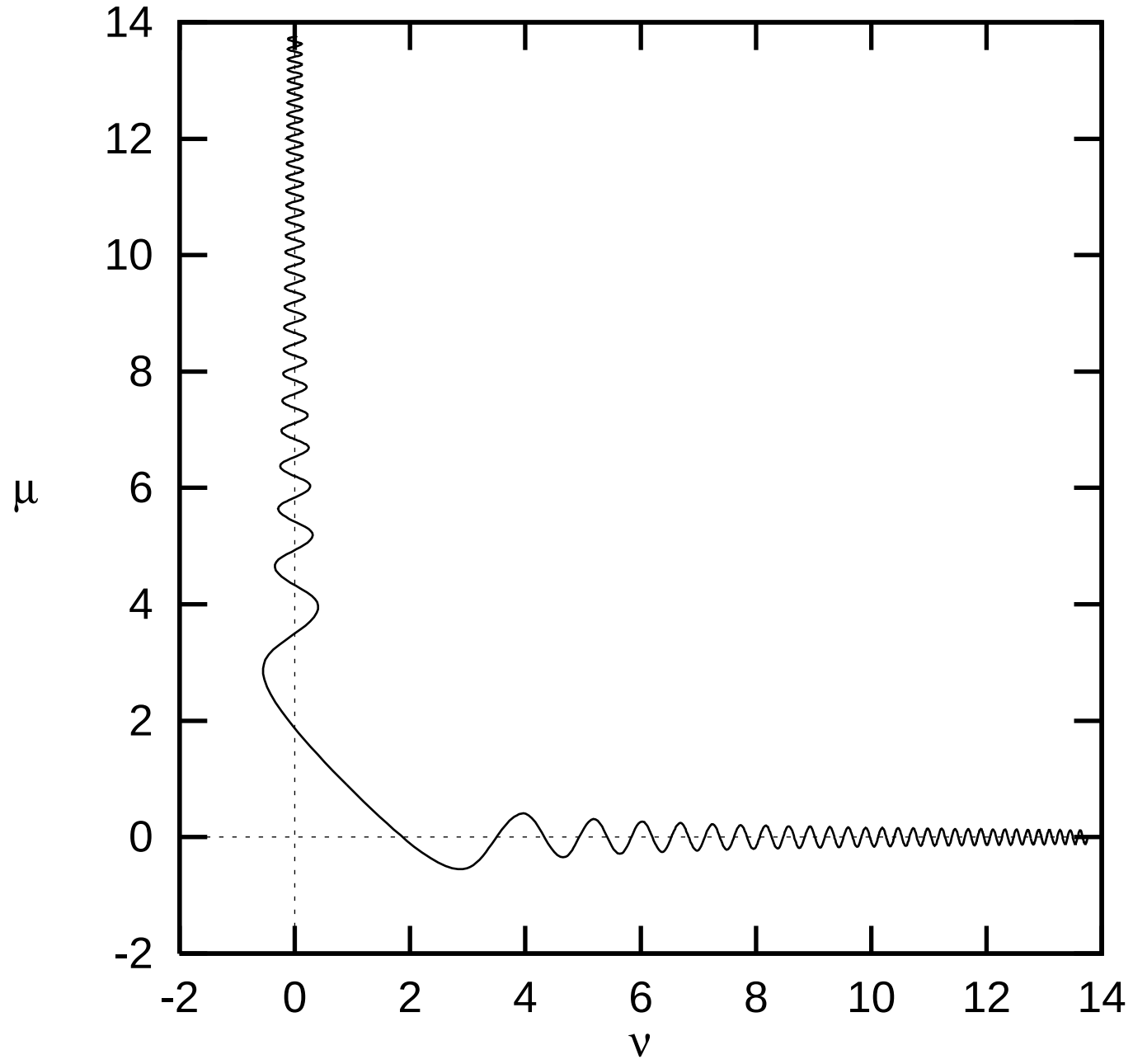

Figure 5: The orbit yielding the escaping trajectory in the Poincaré map in Fig. 4, arriving from $\mu = \infty$ and continuing directly to $\nu = \infty$.

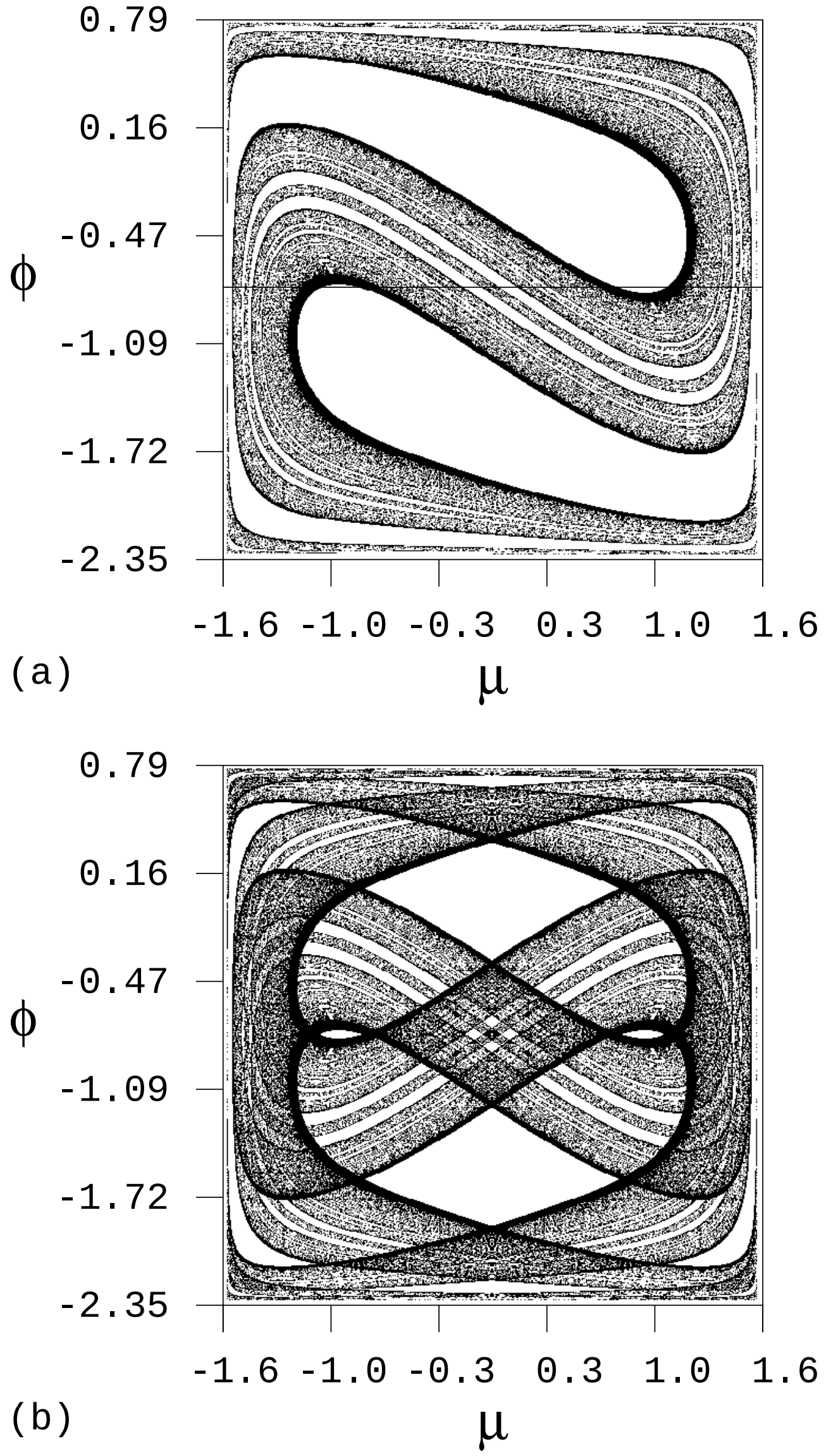

Figure 6: The manifolds of DKP in the Poincaré map $(\mu, \phi)$ for $\nu = \mu$ for $\epsilon = 0.35$. (a) $W^s$, (b) $W^s$ and $W^u$.

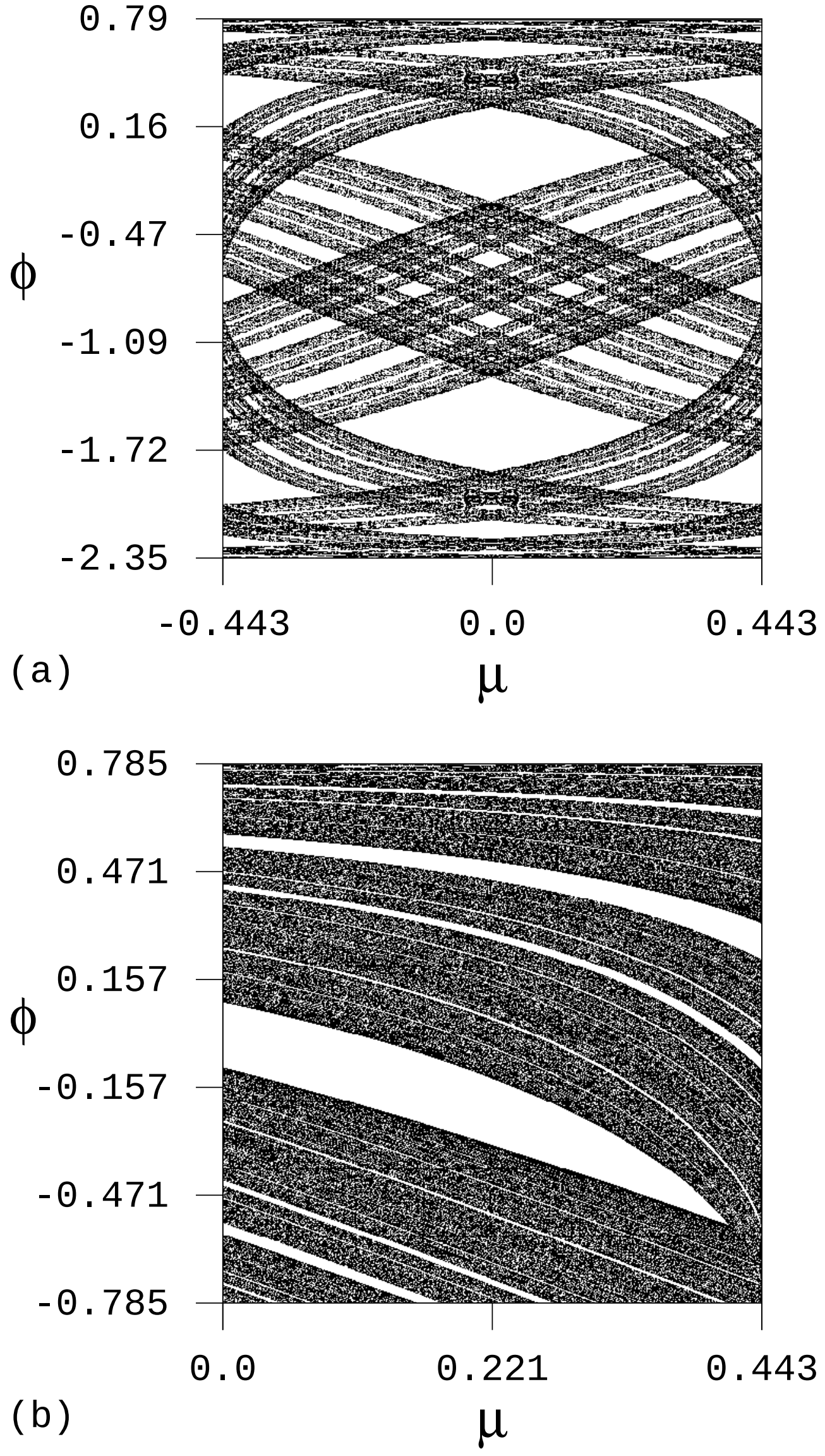

Figure 7: The manifolds of the 4-disk problem. (a) $W^s$ and $W^u$ for $r = 2.3$, (b) $W^s$ for $r = 2.06$.

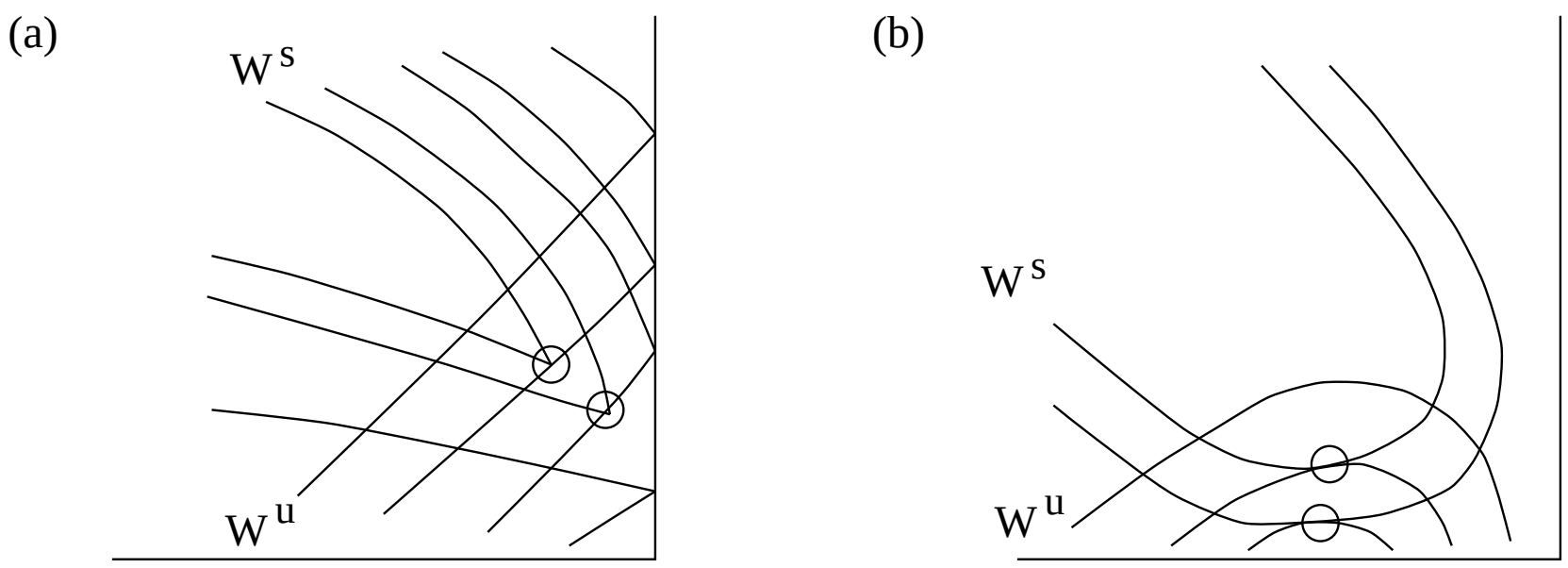

Figure 8: A sketch of the structure of homoclinic tangencies in (a) the 4-disk system, (b) DKP.

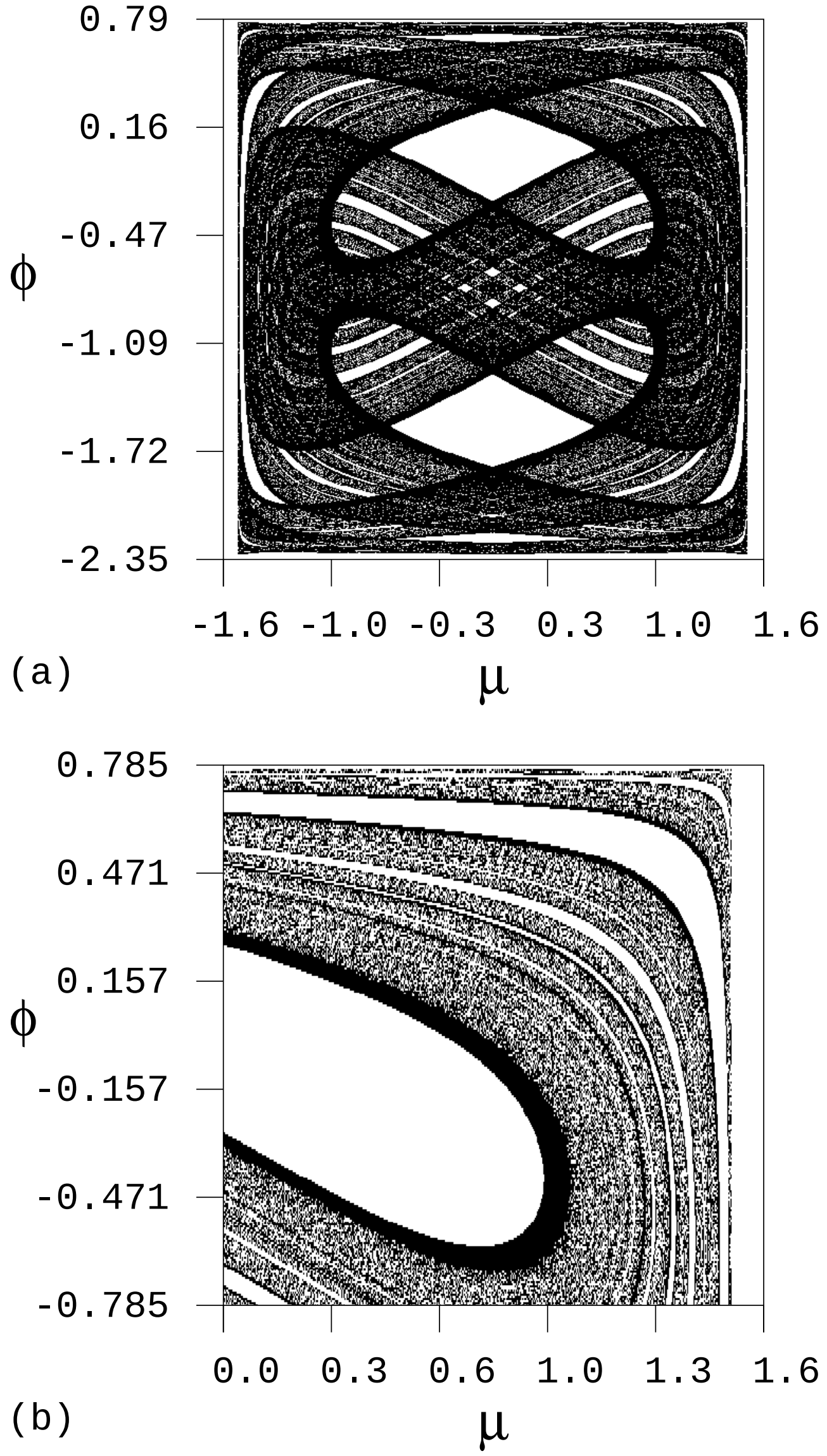

Figure 9: The manifolds of DKP for $\epsilon = 0.2$. (a) $W^s$ and $W^u$ (b) $W^s$.

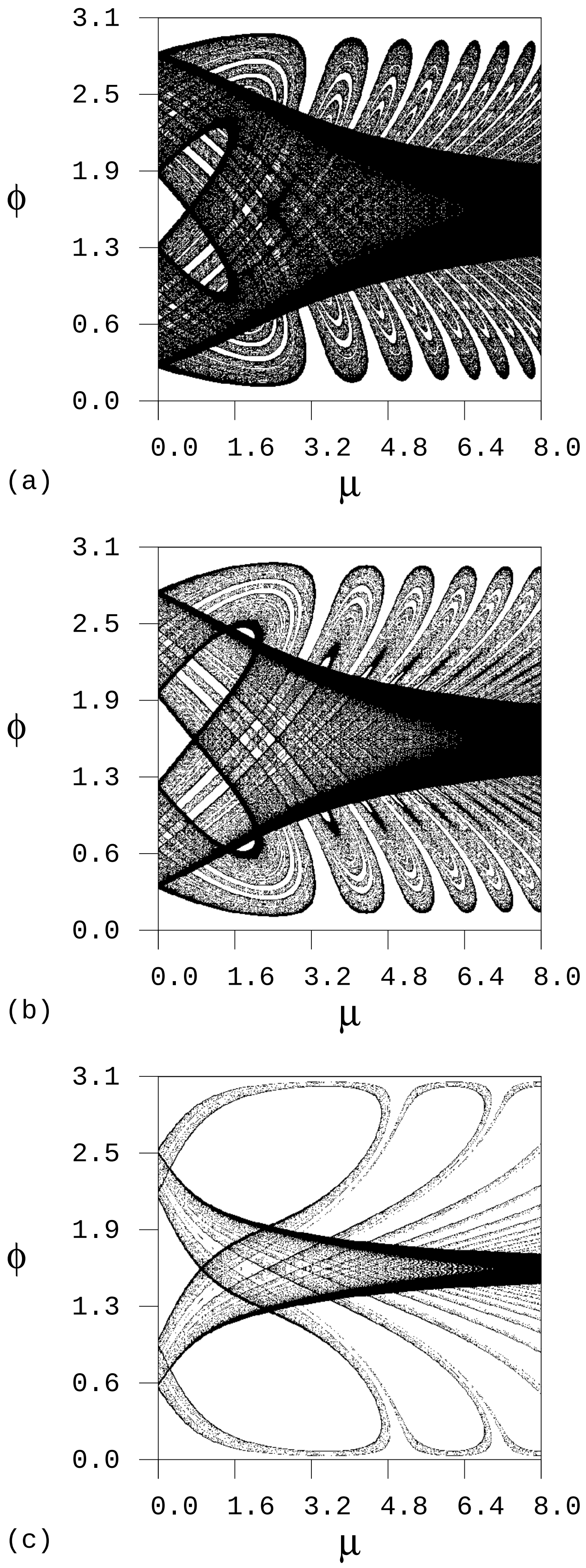

Figure 10: The manifolds of DKP $W^s$ and $W^u$ in the Poincaré map $(\mu, \phi)$ for $\nu = 0$ for (a) $\epsilon = 0.2$, (b) $\epsilon = 0.35$ and (c) $\epsilon = 2.0$

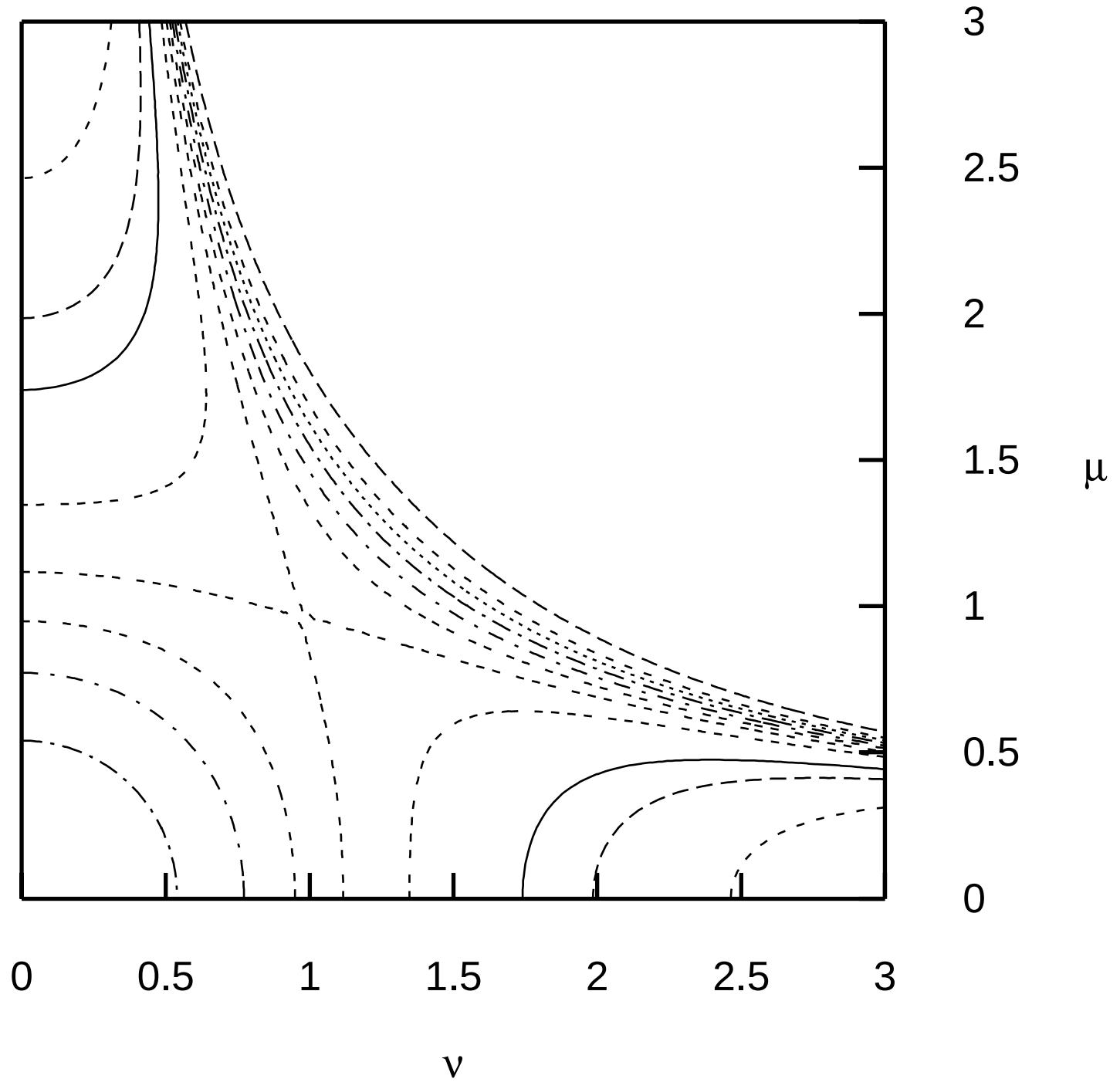

Figure 11: The potential at the critical scaled energy $\epsilon_c = 0.328782$.